\documentclass[a4paper,11pt]{article}

\usepackage{pos}
\usepackage{siunitx}
\usepackage[nohyperlinks, nolist]{acronym}
\usepackage{textgreek}
\usepackage{lineno}

\renewcommand{\logo}{\relax}

\DeclareSIUnit{\h}{h}
\DeclareSIUnit{\hour}{h}
\DeclareSIUnit{\day}{d}
\DeclareSIUnit{\t}{t}

\newcommand{\nuc}[2]{\ensuremath{^{#1}\text{#2}}}

\newcommand{\Qbb}{$Q_{\beta\beta}$}
\newcommand{\nldb}{$0\nu\beta\beta$}

\begin{document}
\title{First operation of poly(ethylene naphthalate) enclosures for high-purity germanium detectors in liquid argon for \nuc{42}{K}/\nuc{42}{Ar} mitigation}
\ShortTitle{PEN enclosures for HPGe detectors to mitigate \nuc{42}{K}}
\author*[a]{Christoph Vogl}
\author[a]{Tommaso Comellato}
\author[a]{Konstantin Gusev}
\author[b,c]{Brennan Hackett}
\author[a,1]{Patrick Krause}
\author[a]{Andreas Leonhardt}
\author[b]{Béla Majorovits}
\author[a]{Niko N. P. N. Lay}
\author[a]{Moritz Neuberger}
\author[a]{Nadezda Rumyantseva}
\author[a]{Mario Schwarz}
\author[a]{Michael Willers}
\author[a]{Stefan Schönert}
\affiliation[a]{Department of Physics, TUM School of Natural Sciences, Technical University of Munich,\\ James-Franck-Str. 1, 85748 Garching, Germany}
\affiliation[b]{Max Planck Institute for Physics,\\ Boltzmannstr. 8, 85748 Garching, Germany}
\affiliation[c]{Oak Ridge National Laboratory, Oak Ridge,\\ TN 37830, USA}
\note{Now at Department of Physics, Simon Fraser University, Burnaby, BC V5A 1S6, Canada}
\emailAdd{christoph.vogl@tum.de}
\abstract{Commercial argon contains cosmogenic \nuc{42}{Ar} whose progeny \nuc{42}{K} is a critical background component for the \acf{legend}.
LEGEND operates \ac{hpge} detectors bare in liquid argon.
\nuc{42}{K} is attracted by the \ac{hpge} detectors' electric fields, and drifts toward the germanium surface, where it undergoes beta decay.
LEGEND-1000 will mitigate \nuc{42}{K}-induced background by using underground-sourced argon, depleted in cosmogenic isotopes.
If underground argon is not available,  mitigation techniques must be employed.
\ac{pen} enclosures were proposed to hinder the ion drift, decrease the beta-particle's energy, and produce scintillation light. In this paper, we report on operating two \ac{hpge} detectors, both bare and \ac{pen}-enclosed, in \nuc{42}{Ar}-enriched liquid argon, and find no evidence for deterioration of energy stability or resolution due to the enclosures. We monitor the beta and gamma rates of \nuc{42}{K}, find complex time-dependencies extending to roughly \qty{30}{days} after applying the \ac{hpge} detectors' high-voltage, and qualitatively demonstrate the \nuc{42}{K} suppression capabilities of enclosures.}
\FullConference{XIX International Conference on Topics in Astroparticle and Underground Physics (TAUP2025)\\
Xichang\\
25.08.2025--29.08.2025
}
\maketitle

\section{Introduction}

LEGEND is searching for neutrinoless double-beta (\nldb) decay of \nuc{76}{Ge} by operating isotopically enriched \acf{hpge} detectors bare in \acf{lar}.
The \ac{lar} acts as a detector coolant as well as passive and active shielding from environmental radio-background.
The current phase of the experiment, LEGEND-200~\cite{legendcollaborationFirstResultsSearch2025}, is taking data in hall A of the \acf{lngs}, while its next phase, LEGEND-1000~\cite{legendcollaborationLEGEND1000PreconceptualDesign2021a}, is  being prepared in hall C of the same laboratory.

\begin{figure}[b]
    \centering
    \includegraphics[scale=0.8]{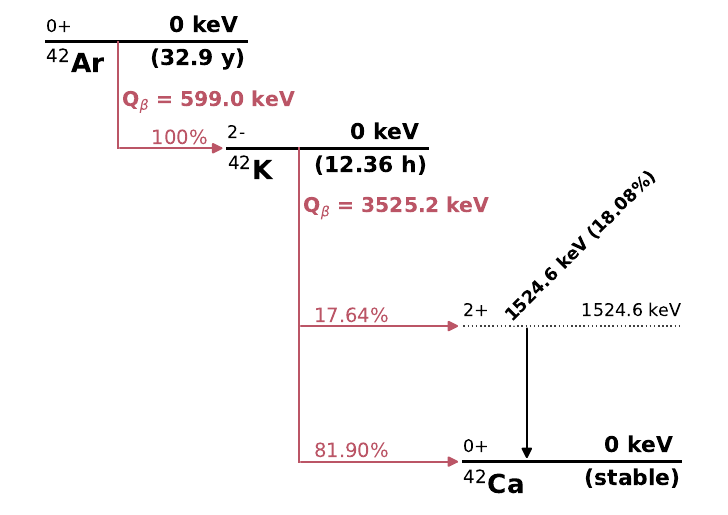}
    \caption{Simplified decay scheme of \nuc{42}{Ar} and \nuc{42}{K}. \nuc{42}{K} is mainly decaying directly to the ground state of \nuc{42}{Ca} and has a higher $Q_\beta$-value than \nuc{76}{Ge}. Therefore, it contributes to background in the region of interest for LEGEND.}
    \label{fig:ar42-k42-decay-scheme}
\end{figure}

Commercial argon is sourced from the atmosphere and contains cosmogenically produced \nuc{42}{Ar}, a beta-unstable isotope decaying into the ground state of \nuc{42}{K} with a long half-life of \qty{32.9}{y} and low $Q_\beta$-value of \qty{599.0}{\keV}.
\autoref{fig:ar42-k42-decay-scheme} show a simplified decay scheme.
\nuc{42}{K} also undergoes beta-decay, with a half-life of \qty{12.36}{\h} and a $Q_\beta$-value of \qty{3525.2}{\keV}.
Its high $Q_\beta$ value surpasses the \Qbb{} value of \nuc{76}{Ge} located at \qty{2039}{\keV} and can therefore populate the signal region of \nldb{} decay with background radiation.
Albeit most \nuc{42}{K} decays end up in the ground state of \nuc{42}{Ca}, in \qty{17.6}{\percent} of the cases, a characteristic gamma line is emitted at \SI{1524.6}{\keV}.
\nuc{42}{K} is produced in a charged state, or forms charged complexes, and is attracted by the \ac{hpge} detectors' electric field~\cite{krzysztofpelczarBackgroundsGERDADARKSIDE2015, lubashevskiyMitigation42Ar2018}.
In GERDA Phase I, \nuc{42}{K} was mitigated by a copper barrier, which screened the electric field and stopped the ion drift.
GERDA Phase II instrumented the \ac{lar} and used transparent \acp{nms} coated with \ac{tpb} to hinder \nuc{42}{K} drift as well as pulse-shape discrimination~\cite{lubashevskiyMitigation42Ar2018}.
LEGEND-1000 plans to use argon sourced from deep underground wells where cosmic rays cannot reach and short-lived radioactive isotopes have decayed long past~\cite{darksidecollaborationResultsFirstUse2016}.

More powerful mitigation techniques must be developed to operate LEGEND-1000 in \ac{atlar} in case \ac{uglar} is unavailable.
Developing and benchmarking transparent and scintillating plastic enclosures for \ac{hpge} detectors is part of LEGEND-1000's \nuc{42}{K} mitigation R\&D plan.
\nuc{42}{K} drifting towards an enclosed detector is stopped on the surface of the enclosure and decays there.
Electrons originating from such decays and depositing signals in the \ac{hpge} detector must traverse the scintillator, where they lose energy, leading to two effects.
First, the electrons reach the \ac{hpge} detector's active volume with less energy, reducing their rate at \Qbb.
Second, scintillation light generated in the enclosure can be detected by the \ac{lar} instrumentation, and the corresponding event can be removed from analysis.
\acf{pen} has emerged as a radiopure, transparent, wavelength-shifting, and scintillating plastic, and is in use in LEGEND-200 as \ac{hpge} detector holder material~\cite{efremenkoProductionValidationScintillating2022, manzanillasOpticalPropertiesLow2022, Manzanillas_2022}.
It is therefore natural to choose \ac{pen} as the first material to investigate when testing scintillating enclosures.

In this paper, we report on the performance of two \ac{pen}-enclosed \ac{hpge} detectors operated in \ac{lar}, as well as on measurements of the observed time dependence of beta and gamma radiation emitted by \nuc{42}{K} with and without enclosures.
The \nuc{42}{K} suppression provided by enclosures will be reported elsewhere.

\section{Experimental setup and data processing}

\begin{figure}[bp]
    \centering
    \includegraphics[height=0.4\linewidth]{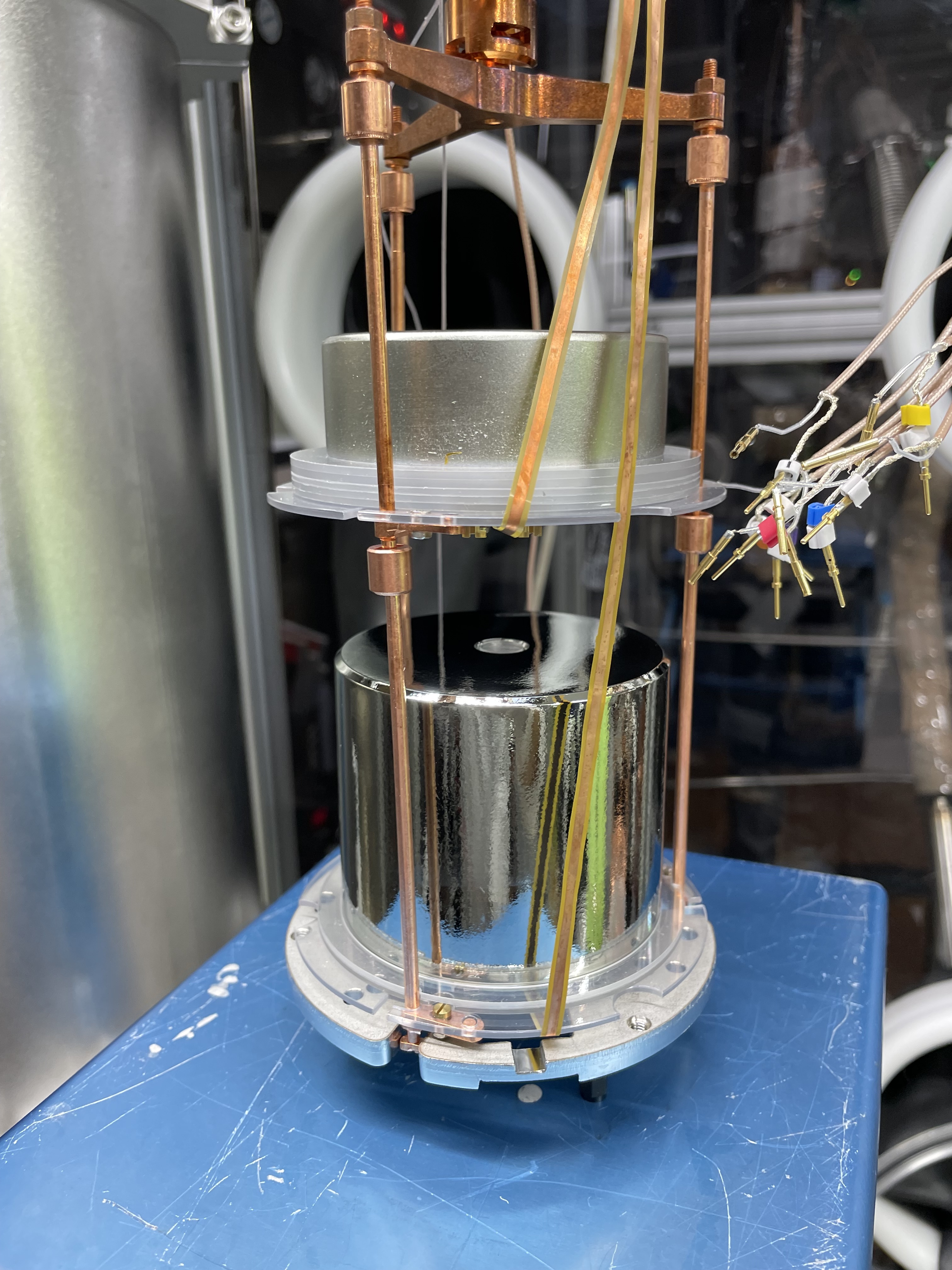}
    \hspace{1cm}
    \includegraphics[height=0.4\linewidth]{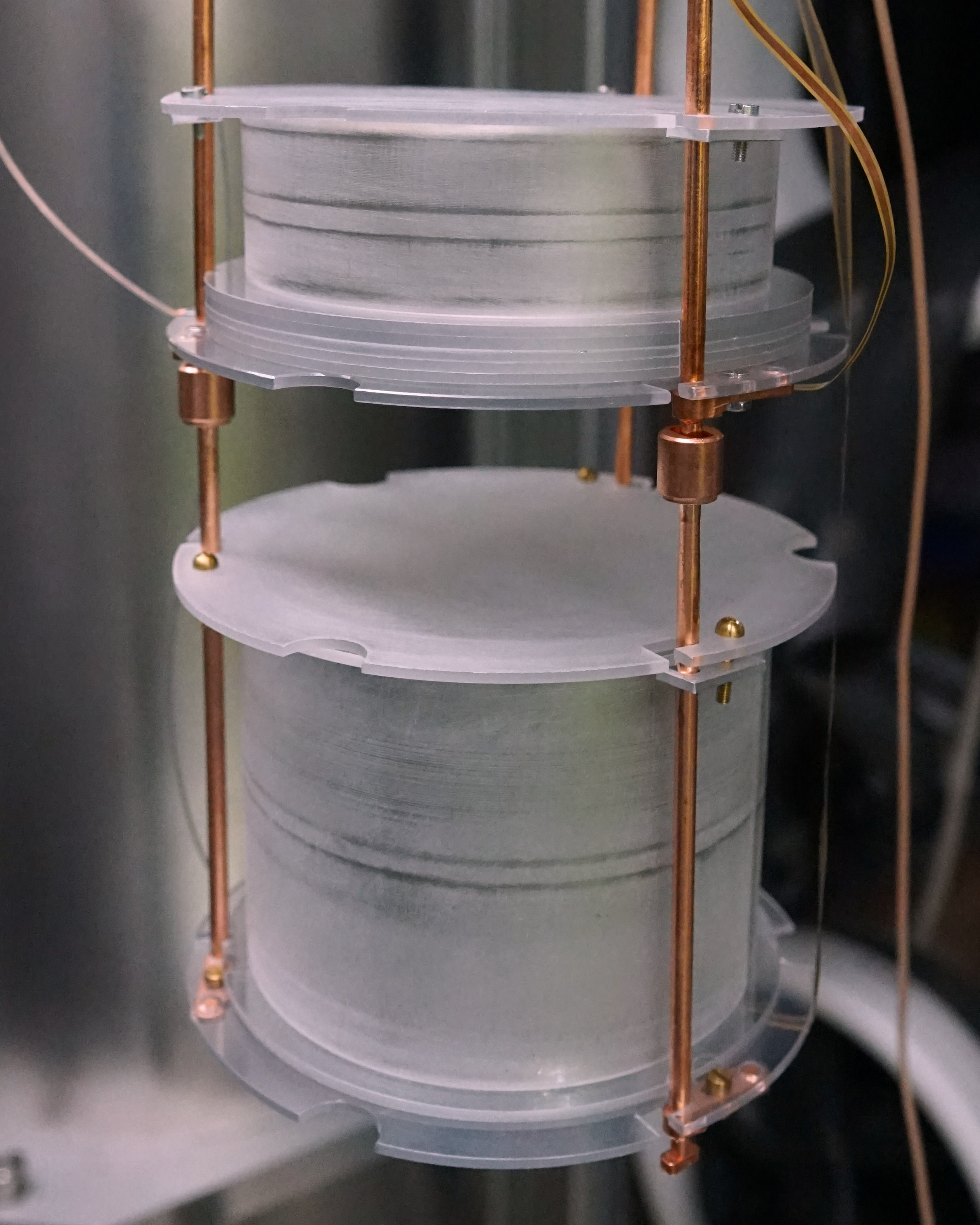}
    \caption{HPGe detectors mounted in a string. A BEGe detector is mounted on top, and an IC detector is mounted on the bottom. The setup is bare on the left and enclosed in PEN on the right.}
    \label{fig:hpge-mounted}
\end{figure}

We mounted two \ac{hpge} detectors of different geometry (\ac{bege} and \ac{ic}) on \ac{pen} plates in a string in two configurations shown in~\autoref{fig:hpge-mounted}.
The bottom plates differ from what is currently used in LEGEND-200; they  feature cut-outs only where necessary for establishing electrical contact.
The \emph{bare} configuration is shown on the left side of the figure, the \emph{enclosed} configuration on the right.
The detectors feature bevels on their lower end, which are filled by \ac{pen} discs of varying inner diameter in both configurations to minimize the area exposed to \nuc{42}{K}.
In the enclosed configuration, a cylinder wrapped around the detector's side and a top plate further reduce the impact of \nuc{42}{K}.
The setup is surrounded by wavelength-shifting fibers coupled to \acp{sipm} (not shown).
The \ac{hpge} detector signals are amplified by close-by cryogenic charge-sensitive amplifiers, while the \ac{sipm} signals are amplified at room temperature.
The payload is lowered through a lock system into the \ac{scarf}~\cite{wiesingerTUMLiquidArgon2014}, a \qty{1}{\t} \ac{lar} cryostat for R\&D related to GERDA and LEGEND.

Data was taken in both configurations, and in pure \ac{lar} and \nuc{42}{Ar}-spiked \ac{lar}.
The production of \nuc{42}{Ar}, and its injection into \ac{scarf} has already been presented in~\cite{schwarz42ArProductionInjection2025}.
In total, \qty[separate-uncertainty=true]{434(8)}{\becquerel} \nuc{42}{Ar} were added to the gas phase of SCARF.
A \emph{background} run, before \nuc{42}{Ar} injection, was taken in the enclosed configuration, and regular \nuc{228}{Th} calibrations were taken for all periods. 
A period (e.g., p05) is defined as a set of runs in the same configuration and may contain physics and calibration runs (e.g., r057, r061).

The \ac{daq} triggers on signals from \ac{hpge} detectors, subsequently reading out all detectors (\ac{hpge} and \ac{sipm}), with raw waveforms stored to disk for offline analysis.
Waveforms used in this analysis have a length of \qty{160}{\micro \s}, centered around the trigger, with a sampling rate of \qty{25}{\mega \Hz}.
\ac{dsp} is implemented with \texttt{dspeed}~\cite{guinnDspeed2025}, and energy calibration routines from \texttt{pygama}~\cite{agostiniPygama2024} are used.
We employ a standard \ac{dsp} composed of pole-zero correction using a separate parameter per \ac{hpge} detector and period, and subsequent trapezoidal filtering with fixed parameters (rise time: \qty{7}{\micro \s}, flat top: \qty{3}{\micro \s}).
The energy is picked up at \qty{80}{\percent} into the flat top.
An optimized \ac{dsp} is also used to address possible changes in noise conditions between runs. 
It features a double-pole-zero correction and a trapezoidal filter with parameters optimized for each run.
The time constant of the first pole is identical to the one in the standard \ac{dsp}, and the energy is picked up at the same location.
We apply data cleaning to remove unstable time frames, quality cuts on \ac{dsp} parameters to eliminate noisy and unphysical waveforms, and the \ac{lar} instrumentation to remove muon-correlated events.
Pulse-shape discrimination~\cite{agostiniPulseShapeAnalysis2022} is applied to the \ac{ic} detector to remove alpha events originating from implanted \nuc{214}{Pb}/\nuc{210}{Po}~\cite{tommasocomellatoInvertedCoaxialDetectors2022}.

\section{Energy resolution and stability}

\begin{figure}[htbp]
    \centering
    \includegraphics[scale=1]{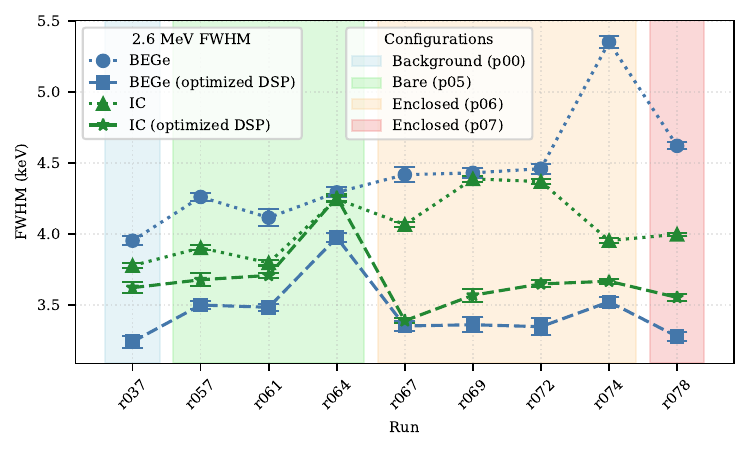}
    \caption{Energy resolution at \qty{2.6}{\MeV} of all \nuc{228}{Th} calibration runs in periods 0, 5, 6, and 7. No evidence was found to suggest that \ac{pen} enclosures deteriorate the \ac{hpge} detectors' energy resolutions.}
    \label{fig:energy-resolutions}
\end{figure}

The energy resolutions of the \ac{hpge} detectors, measured by the \ac{fwhm} of the \qty{2615}{\keV} peak, are plotted in \autoref{fig:energy-resolutions} for all \nuc{228}{Th} calibration runs of the periods of interest in this analysis.
The energy resolutions of runs performed in the enclosed configuration (p06 and p07) are slightly higher for both detectors than for runs in the bare configuration (p05) using the standard \ac{dsp}.
The \ac{bege} detector has a larger than usual \ac{fwhm} in r074.
Both of these effects can likely be attributed to different noise conditions, as becomes apparent when comparing with the optimized \ac{dsp} energy resolution values.
There, the outlier at r074 vanishes, and p06 and p07 feature an energy resolution at least as good as p05.
No evidence of a deterioration of energy resolution due to \ac{pen} enclosures was found.

The bottom panel of \autoref{fig:beta-gamma-time-dependence} shows the time stability of the \qty{1524.6}{\keV} peak in the \ac{ic} detector for all physics runs in p05, p06, and p07.
The data are interrupted due to calibration runs, hardware changes, and environmental changes, which deteriorate data quality.
The \qty{1524.6}{\keV} peak position and width are stable throughout the dataset.
No evidence was found to suggest that \ac{pen} enclosures harm operational stability.
The bottom right panel shows a projection onto the energy axis, displaying an unbiased peak of regular shape.

\section{Time evolution of \texorpdfstring{\nuc{42}{K}}{K-42} gamma and beta rates}

\begin{figure}[htbp]
    \centering
    \includegraphics[scale=1]{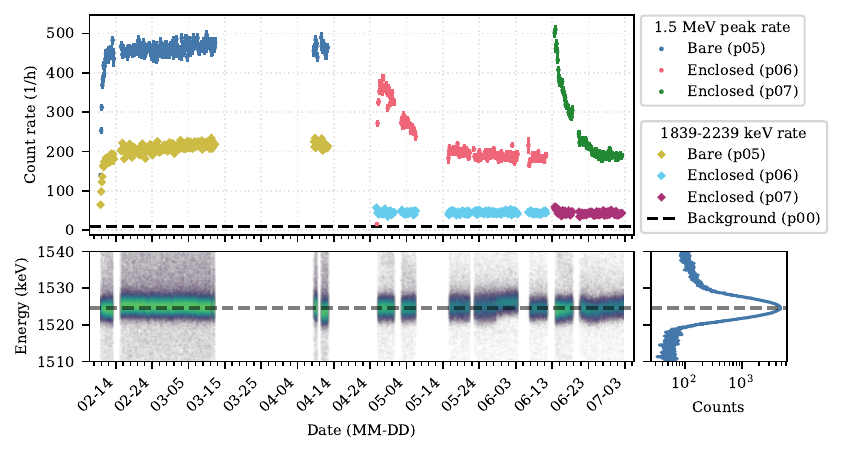}
    \caption{Time evolution of beta and gamma radiation by \nuc{42}{K} recorded with the IC detector (top) for all physics runs in p05, p06, and p07. The rates show a complex time-dependence.
    Partially filled bins are corrected.
    Energy stability of the \qty{1524.6}{\keV} gamma line (bottom) for the same dataset, along a projection onto the energy axis (bottom right). The peak position and width are approximately stable in time. The standard \ac{dsp} is used.}
    \label{fig:beta-gamma-time-dependence}
\end{figure}

The electric field of \ac{hpge} detectors influences the drift and spatial distribution of \nuc{42}{K}; hence, we expect gamma and beta rates to vary in time after the detectors have been biased.
The spatial distribution of \nuc{42}{K} affects the rates differently: (i) \qty{1524.6}{\keV} gammas have an attenuation length of around \qty{15}{\cm} in \ac{lar}~\cite{seltzerXCOMPhotonCrossSections1987}, thus moderately remote decays contribute, while (ii) \nuc{42}{K} betas depositing energy around \Qbb{} probe only on-surface or close-by \nuc{42}{K} activities.

The top panel of \autoref{fig:beta-gamma-time-dependence} shows the time-resolved evolution of gamma (from the sideband-subtracted \qty{1524.6}{\keV} peak) and beta (from a \qty{400}{\keV}-wide window around \Qbb) rates for p05 to p07, using the \ac{ic} detector\footnote{The discussion and findings in the following apply to the \ac{bege} detector as well.}.
The dashed line in the beta region represents the environmental background and is extracted from p00, i.e., before the injection of \nuc{42}{Ar}.

At the beginning of p05 (bare), when the detectors have been recently biased, the gamma rate increases, due to \nuc{42}{K} being attracted towards the \ac{hpge} detectors.
After a couple of days, the increase slows down, and following a gap in usable physics data lasting around \qty{25}{days}, the gamma rate reappears at roughly the same value as before, suggesting that equilibrium has been reached.
The beta rate follows the gamma rate, indicating that the concentration of close-by \nuc{42}{K} increases due to a global motion towards the \ac{hpge} detectors.

Between p05 and p06, the \ac{hpge} detectors were switched off, lifted, enclosed in \ac{pen}, and re-deployed.
They were re-biased shortly before commencing p06 data-taking.
In the first \qty{24}{hours} of p06, the gamma line intensity increases rapidly before it drops and levels off around \qtyrange{20}{30}{days} after the maximum.
We interpret this behavior as \ac{pen} charging up on a timescale of days, increasing the mean distance of the \nuc{42}{K} distribution to the detectors.
In contrast to p05, the beta rate is decoupled from the gamma rate and remains constant, i.e., close-by or on-surface concentrations remain unaffected by the longer-range ion drift or charge-up.
The beta rate being much lower than in p05 demonstrates the potential of enclosures for \nuc{42}{Ar}/\nuc{42}{K} mitigation.

To confirm the measured effects, the detectors were unbiased for two days between p06 and p07, and then re-biased; no further changes were made.
Surface charges are expected to diminish.
The data in p07 exhibits a qualitatively similar behavior to that in p06.
A detailed understanding of the \nuc{42}{K} spatial distribution and its time-dependence is beyond the scope of this work, as it requires extensive simulations of the entire setup, including the cryostat, and accounting for electric fields and ion drifts.

\section{Conclusion}

We operated two \ac{hpge} detectors in \ac{lar} in two configurations: (i) bare for more than a month, and (ii) enclosed in \ac{pen} for over two months.
Comparing the data taken with \ac{pen} enclosures to the bare configuration, no evidence was found for deterioration in energy stability or resolution.

We observed gamma and beta radiation from \nuc{42}{K} decays in an \nuc{42}{Ar}-spiked \ac{lar} environment and noted non-trivial time dependencies for the enclosed case. 
The intensity of the \qty{1524.6}{\keV} line of \nuc{42}{K} increases rapidly after the enclosed detectors are biased, and subsequently decreases at a slower rate.
The beta rate remains constant, however.
\ac{pen} enclosures were found to significantly reduce the count rate of \nuc{42}{K} betas around \Qbb{} of \nuc{76}{Ge}.
A quantitative analysis of the \nuc{42}{K} survival fraction provided by enclosures is outside the scope of this paper, and will be reported elsewhere.

\acknowledgments
This work is supported by the Deutsche Forschungsgemeinschaft (DFG, German Research Foundation) through the Excellence Cluster ORIGINS (EXC 2094-39078331) and the Collaborative Research Center SFB1258 ''Neutrinos and Dark Matter in Astro- and Particle Physics`` (SFB1258-283604770).

\begin{acronym}
    \acro{legend}[LEGEND]{Large Enriched Germanium Experiment for Neutrinoless \textbeta\textbeta{} Decay}
    \acro{lar}[LAr]{Liquid Argon}
    \acro{uglar}[UGLAr]{Underground Liquid Argon}
    \acro{atlar}[AtLAr]{Atmospheric Liquid Argon}
    \acro{hpge}[HPGe]{High-Purity Germanium}
    \acro{bege}[BEGe]{Broad Energy Germanium}
    \acro{ic}[IC]{Inverted Coaxial}
    \acro{sipm}[SiPM]{Silicon Photomultiplier}
    \acro{nms}[NMS]{Nylon Mini-Shroud}
    \acro{tpb}[TPB]{Tetraphenyl Butadiene}
    \acro{pen}[PEN]{Poly(ethylene naphthalate)}
    \acro{lngs}[LNGS]{Laboratori Nazionali del Gran Sasso}
    \acro{scarf}[SCARF]{Subterranean Cryogenic Argon Research Facility}
    \acro{large}[LArGe]{Liquid Argon Germanium}
    \acro{daq}[DAQ]{Data Acquisition}
    \acro{dsp}[DSP]{Digital Signal Processing}
    \acro{fwhm}[FWHM]{Full Width at Half Maximum}
\end{acronym}

\bibliographystyle{JHEP}
\bibliography{TAUP2025_proceeding}
\end{document}